\def\aa{{A\&A}}

\def\annrev{{ARA\&A}}
\def\apj{{ApJ}}

\documentclass[cup5b]{caps}
\usepackage{graphicx}
\usepackage{amssymb}
\usepackage{ociwsymp3e}  

\HeadText{A. K. Romer et al.}  

\begin{document}

\pagenumbering{arabic}


%

\author[]{A. K. Romer$^{1}$, P.L. Gomez$^{1}$, and the 
ACBAR collaboration$^{2,3,4,5,6,7,8}$\\ (1) Carnegie Mellon
University, (2) Cardiff University, (3) Jet Propulsion Laboratory,\\
(4) Lawrence Berkeley National Laboratory, (5) University of
California at Berkeley,\\ (6) Case Western Reserve University (7)
University of California, Santa Barbara,\\ (8) California Institute of
Technology }

\chapter{Imaging the Sunyaev Zel'dovich Effect using ACBAR on Viper}

\begin{abstract}

During 2001 and 2002, observations of several X-ray bright clusters of
galaxies were conducted using the ACBAR Bolometer Array on the South
Pole Viper telescope.  A multi-frequency analysis of these clusters is
currently underway. This multi-frequency analysis includes 150, 220
and 275 GHz data from ACBAR, X-ray imaging and spectroscopy from
Chandra and XMM-Newton, and weak lensing data from the CTIO 4m Blanco
telescope. We describe here how ACBAR can be used to create fully
sampled cluster images and present such images for four of the
clusters in our sample; Abell 3266, Abell 3827, Abell S1063 and
1E0657-56. In these images, the Sunyaev-Zel'dovich Effect is clearly
detected.

\end{abstract}

\section{Introduction}

The Arcminute Cosmology Bolometer Array Reciever instrument (ACBAR;
Runyan et al. 2003) has been operating on the 2.1m
millimeter/submillimeter Viper telescope since January 2001. The Viper
telescope (Peterson et al. 2000) is located at one of the best sites
in the world for millimeter/submillimeter astronomy (Peterson et
al. 2002), the South Pole Martin A. Pomeranz Observatory. The ACBAR
focal plane is tiled with 16 microlithographed ``spider-web''
bolometers arranged in a $4\times4$ array. Each bolometer has a
Gaussian beam with a $\simeq4.5'$ FWHM (see Figures 2 and 12 in Runyan
et al. 2003). Adjacent beams are separated on the sky by $\simeq
15'$. During the 2001 observing season, the array was arranged so that
each column of four bolometers was sensitive to one of the following
mean frequencies; 150, 220, 275 and 345 GHz. During the 2002 season,
the 345 GHz bolometers were replaced by 150GHz bolometers. The array
was also re-arranged so that now each row, rather than each column,
was sensitive to the same frequency. Two rows of 150GHz bolometers,
and one row each of 220 and 275 GHz bolometers, were used. The ACBAR
observing frequencies straddle the peak in the CMB spectrum and the
220 GHz ``null'' point in the thermal Sunyaev-Zel'dovich effect
(Sunyaev \& Zel'dovich 1972; SZE hereafter). The primary calibrators
for the 2001 and 2002 seasons were Mars and Venus respectively. During
the periods when these planets were not accessible from the South
Pole, various Galactic sources were used to monitor variations in the
calibration. The total calibration uncertainty is estimated to be 10\%
(Runyan et al. 2003).  The Galactic source observations were also used
to determine the telescope pointing model. The overall pointing error
for the 2001 season was $\simeq1'.4$ in right ascension and
$\simeq20''$ in declination. For the 2002 season, a mechanical chopper
encoder was replaced with an optical encoder and the pointing error in
the right ascension direction improved to $\simeq 20''$.

To date, ACBAR data have been applied to three different
projects. First, to determine the Cosmic Microwave Background (CMB)
power spectrum over the range $150<l<3000$ with a resolution of
$\Delta l=150$ (Kuo et al. 2003).  This spectrum has yielded
constraints on a variety of cosmological parameters (Goldstein et
al. 2003, Spergel et al. 2003). Second, we are using the 150 GHz maps
generated for the CMB anisotropy project to search for high redshift
clusters of galaxies via the SZE (Runyan 2002). Several cluster
candidates have been selected using a matched spatial filtering
technique. These candidates are being followed up with optical and
X-ray imaging to confirm their identification and to measure
redshifts. Ultimately, the high redshift, or ``blind'', ACBAR cluster
survey will be used to derive complimentary (to the CMB power
spectrum) constraints on cosmological parameters. Finally, we are
using ACBAR, in conjunction with data from X-ray and optical
telescopes, to study a small sample of well known, X-ray bright,
clusters. We report on the recent progress of this cluster study
below.

\section{The Viper Sunyaev-Zel'dovich Survey}

The Viper Sunyaev-Zel'dovich Survey (VSZS) aims to collect SZE, X-ray
and weak lensing data for a complete sample of X-ray clusters. The
aims of the VSZS include a thorough investigation into systematic
biases associated with the computation of the Hubble Constant from SZE
measurements. We will also explore the physical origin of small scale
X-ray surface brightness and/or temperature fluctuations in the
clusters. By comparing ACBAR and X-ray data in those regions, we
should be able to distinguish between contact discontinuities and
shocks in the intracluster gas. Moreover, with access to high quality
X-ray, SZE and weak lensing data, we can measure masses (both total
and baryonic) for the VSZS clusters in a variety of ways. We can
therefore predict the accuracy to which future SZE or X-ray surveys
will be able to measure cluster masses from a single observable such
as SZE signal or X-ray luminosity. These surveys will be based on new
X-ray data ({\it e.g.} the XMM Cluster Survey, Romer et al. 2001) and
new SZE facilities ({\it e.g.} the ACBAR Blind Survey, Runyan 2002;
the Planck Satellite, White 2003; the SZ Array and the South Pole 8m
telescope, Carlstrom et al.. 2003). Their success will be determined
by their ability to measure the evolution in the cluster number
density as a function of halo mass.

All our scientific goals require that we remove signals due to CMB
primary anisotropies from our cluster images. For the low redshift
clusters in our sample (see below for a list) we are able to use the
three ACBAR observing frequencies (150, 220, 275 GHz) to remove the
CMB. Intuitively, this method is very simple as it can be shown that
at 150 GHz we detect the CMB minus the cluster thermal SZE signal, at
220 GHz we detect just the CMB, and at 275 GHz we detect the CMB plus
the cluster signal. In practice, we construct a linear combination of
the 150, 220 and 275 GHz that maximizes the contrast of the SZE over
the CMB and noise. For the higher redshift clusters in our sample, we can use
spatial filtering to make CMB free maps at both 150 and 275GHz; the
CMB anisotropy power spectrum peaks on degree scales whereas the high
redshift clusters cover only a few square arcminutes (see Figures 4
and 5).

The VSZS cluster sample comprises of two subsamples. The low redshift
subsample contains the seven clusters in the REFLEX cluster catalog
(Bohringer et al. 2001) that are brighter than $L_x = 5\times10^{44}
{\rm erg\,s^{-1}}$ (0.1-2.4 keV), have declinations less than
$\delta=-40^{\circ}$ and redshifts less than $z=0.1$; Abell 3367,
A3827, A3266, A3112, A3158, A3921 and A3911. The high redshift
subsample currently\footnote{We plan to extend this sample to include
lower luminosity REFLEX clusters, ACBAR Blind cluster candidates and
southern MACS (Ebeling et al. 2000) clusters during the 2003 \& 2004
seasons.}  contains the three REFLEX clusters with $z>0.1$,
$\delta<-40^{\circ}$ and $L_x > 20\times10^{44} {\rm erg\,s^{-1}}$;
Abell S1063, 1E0657-56 and AS520. To date, all but A3921 and A3911
have been observed with ACBAR and all but AS1063 and A3911 have been
observed with XMM and/or Chandra. Weak lensing observations from the
Mosaic camera on the CTIO 4m Blanco telescope are also available for
all clusters bar AS1063 and A3112.

\section{Imaging VSZS Clusters with ACBAR}

To build up an image of a cluster using ACBAR one uses a raster scan
technique. The sixteen instrument beams are swept across the sky in
co-elevation (which is essentially co-declination at the South Pole)
using a flat tertiary mirror. The tertiary mirror chops back and forth
by up to 3 degrees in a fraction of a second (0.7 Hz in 2001 and 0.3
Hz in 2002). After a certain number (84 in 2001 and 36 in 2002) of
chopper cycles, the elevation is shifted slightly and the process
repeated. During 2001, we experimented with a variety of different
cluster mapping strategies. For the 2002 cluster observations, we
adopted a single strategy. For each cluster we created square
$1.5^{\circ}\times1.5^{\circ}$ maps by setting the chopper throw to
1.5 degrees and using 90 one arcminute elevation shifts.

As part of every cluster observation we made observations of leading
and trailing fields. These ``lead'' and ``trail'' observations were
interleaved with the ``main'' observations. At any given elevation
setting, the observing time in the lead and trail fields was set to
half of that in the main field. Typically, the main would be observed
for 60 seconds and the lead and trail for 30 seconds each. These
observations were used during the data analysis to remove chopper
synchronous offsets. These offsets originate from non-astronomical
sources of microwave radiation such as the sky, the ground and the
telescope optics.

The data from each bolometer element is stored as a time stream of
voltages. To create maps from these time streams, we first bin the
data by chopper position and then subtract the chopper synchronous
offset. Next, we convert from units of voltage (as a function of
chopper position and elevation) to temperature (as a function of
celestial coordinates) using the instrument calibration and telescope
pointing models. At this stage, certain scans are discarded because
they have anomalously high chopper synchronous offsets. This is an
indication of snow accumulation on the mirrors; snow boosts the offset
signal and attenuate the astronomical signal and makes it very hard to
maintain an accurate calibration.  Finally all the scans at a certain
frequency are projected onto a template grid of the sky so they can be
co-added in a noise weighted fashion to produce a fully sampled image.

\section{ACBAR Cluster Images}

In Figures 1 through 5 we present ACBAR images of four VSZS clusters;
Abell 3266, Abell 3827, Abell S1063 and 1E0657-56. The clusters are
presented in order of increasing redshift, from A3266 at $z=0.05$ to
AS1063 at $z=0.34$. Figure 2 shows a 275 GHz image of A3266. In this
image, the cluster shows up as a hot region, as expected because we
are observing at at frequency above the null in the thermal SZE
spectrum. All the other Figures show 150GHz images. In these, the
cluster appears as a cold region (150GHz lies below the thermal
null). The ACBAR beam size (FWHM) is indicated by a solid white circle
in the bottom right corner of each image. To illustrate how the X-ray
and SZE properties of these clusters compare, we have overlaid ROSAT
X-ray surface brightness contours on the ACBAR images. For A3266,
A3827, 1E067-56 the contours are based on data from the ROSAT pointing
archive; a 13ks PSPC observation, a 13ks HRI observation and a 58ks
HRI pointing respectively.  The contours on Figure 5 represent ROSAT
All-Sky Survey data (ROSAT pointing data are not available for
AS1063).

A more detailed analysis of the SZE and X-ray observations of these,
and other ACBAR, clusters will be presented in Gomez et al 2003 (in
preparation).

\paragraph{Acknowledgments}

The ACBAR collaboration comprises of P.A.R. Ade, J.J. Bock,
C.M. Cantalupo, M.D. Daub, J.H. Goldstein, W.L. Holzapfel, C.L. Kuo,
A.E. Lange, M. Lueker, M. Newcomb, J.B. Peterson, C. Reichardt, J.E. Ruhl,
M.C. Runyan and E. Torbet. We acknowledge the REFLEX collaboration for
providing the redshift for AS1063 pre-publication. We thank William
Chase for assistance with data reduction. AKR and PLG acknowledge
financial support from the NASA LTSA grant NAG5-7926.  Computer
equipment for this project was purchased using two AAS small research
grants. Additional computing resources were kindly made available to
us by the PICA group (picagroup.org). ACBAR has been supported by the
NSF Center for Astrophysics Research in Antartica and by NSF grants
OPP-8920223 and OPP-0091840.


  \begin{figure}
    \centering
    \includegraphics[width=10cm,angle=0]{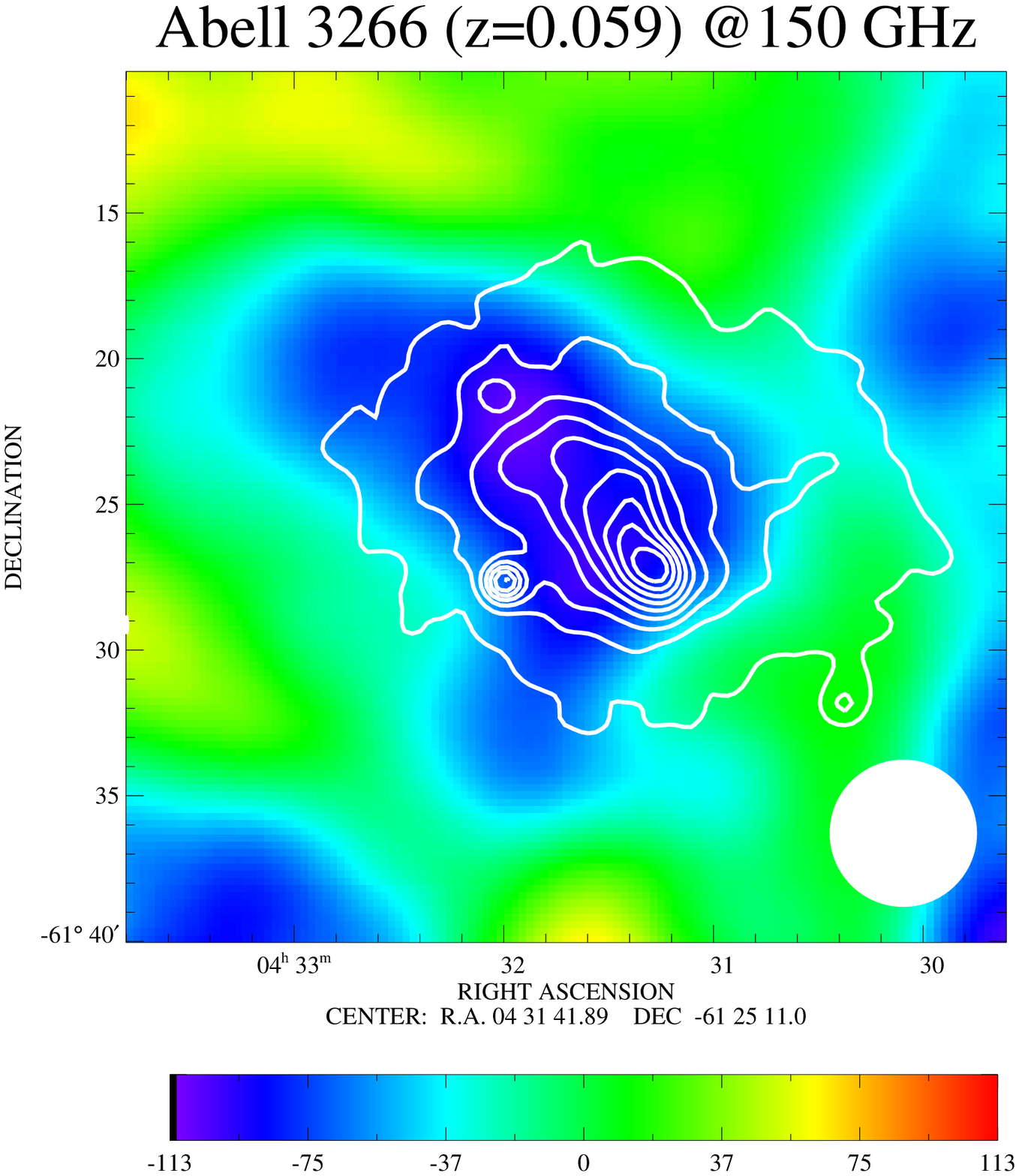}

    \caption{A 150GHz image of A3266 made using ACBAR during 2001. The
    units are $\mu$K. The white contours represent the ROSAT PSPC
    (0.4-2.0 keV) count rate image of A3266. The ACBAR beam size is
    illustrated by the white circle in the bottom right corner of the
    Figure. The noise level in this image is $\simeq$25 $\mu$K rms in
    a $4'\times4'$ region (roughly one beam).}

    \label{A3266_150}
  \end{figure}

  \begin{figure}
    \centering
    \includegraphics[width=10cm,angle=0]{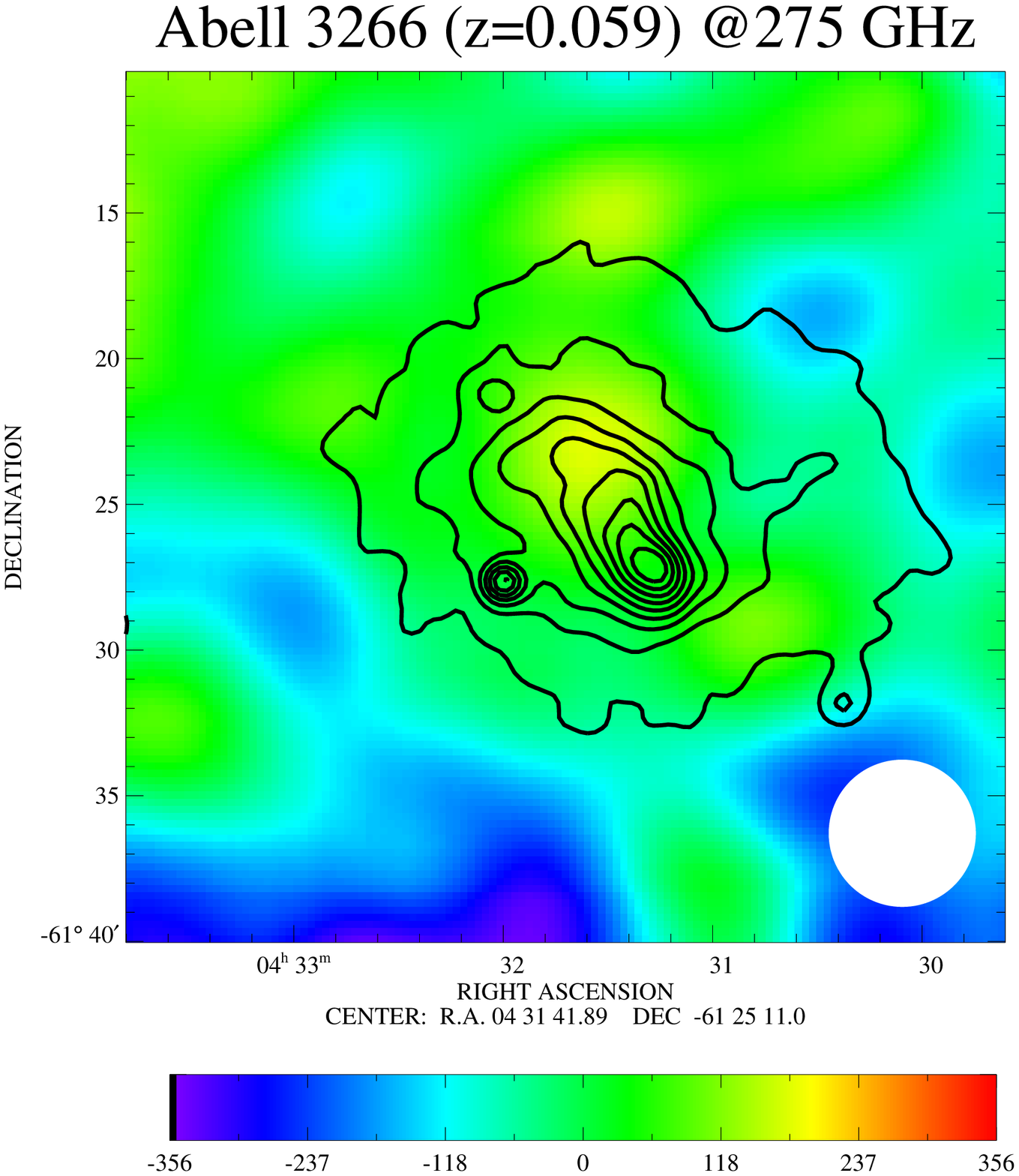}
    
    \caption{A 275GHz image of A3266 made using ACBAR during 2001. The
    units are $\mu$K. The black contours represent the ROSAT PSPC
    (0.4-2.0 keV) count rate image of A3266. The ACBAR beam size is
    illustrated by the white circle in the bottom right corner of the
    Figure. The noise level in this image is $\simeq$70 $\mu$K rms in
    a $4'\times4'$ region (roughly one beam).}

    \label{A3266_275}
  \end{figure}
 
 \begin{figure}
    \centering
    \includegraphics[width=10cm,angle=0]{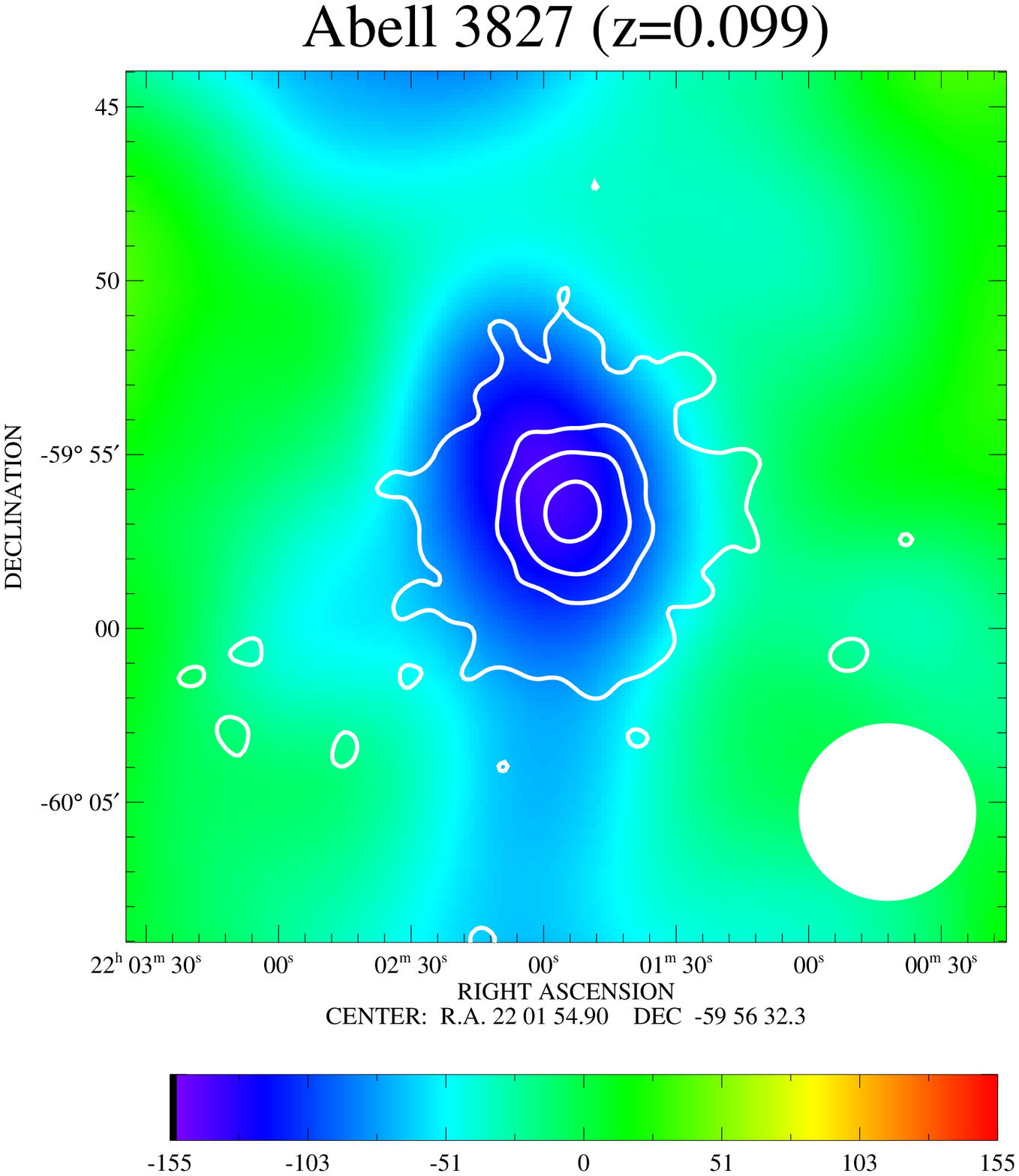}

    \caption{A 150GHz image of A3827 made using ACBAR during 2002. The
    units are $\mu$K. The white contours represent the ROSAT HRI
    (0.1-2.4 keV) count rate image of A3827. The ACBAR beam size is
    illustrated by the white circle in the bottom right corner of the
    Figure. The noise level in this image is $\simeq$24 $\mu$K rms in
    a $4'\times4'$ region (roughly one beam). }

    \label{A3827}
  \end{figure}

  \begin{figure}
    \centering

    \caption{A 150GHz image of 1E0657-56 made using ACBAR during
    2002. The units are $\mu$K. The white contours represent the ROSAT
    HRI (0.1-2.4 keV) count rate image of 1E0657-56. The ACBAR beam
    size is illustrated by the white circle in the bottom right corner
    of the Figure. The noise level in this image is $\simeq$15 $\mu$K
    rms in a $4'\times4'$ region (roughly one beam). }

    \label{1ES}
  \end{figure}

  \begin{figure}
    \centering

    \caption{A 150GHz image of AS1063 made using ACBAR during
    2002. The units are $\mu$K. The white contours represent the hard
    band (0.5-2.0 keV) ROSAT All-Sky Survey count rate image of
    AS1063. The ACBAR beam size is illustrated by the white circle in
    the bottom right corner of the Figure. The noise level in this
    image is $\simeq$31 $\mu$K rms in a $4'\times4'$ region (roughly
    one beam).}

    \label{AS1063}
  \end{figure}


\begin{thereferences}{}

\bibitem{2001A&A...369..826B} B{\" o}hringer, 
H.~et al.\ 2001, \aa, 369, 826 

\bibitem{2002ARA&A..40..643C} Carlstrom, 
J.~E., Holder, G.~P., \& Reese, E.~D.\ 2002, \annrev, 40, 643 

\bibitem{2002astro.ph.12517G} Goldstein, J.~H.~et 
al.\ 2003, \apj~ in press (astro-ph/0212517)


\bibitem{2002astro.ph.12289K} Kuo, C.~L.~et al.\ 2003, \apj~ in press
(astro-ph/0212289)

\bibitem{2000ApJ...532L..83P} Peterson, J.\ B.\ et 
al.\ 2000, \apj~, 532, L83

\bibitem{2002astro.ph.11134P} Peterson, J.~B., 
Radford, S.~J.~E., Ade, P.~A.~R., Chamberlin, R.~A., O'Kelly, M.~J.,
Peterson, K.~M., \& Schartman, E.\ 2003, \apj~ in press
(astro-ph/0211134)

\bibitem{2001ApJ...547..594R} Romer, 
A.~K., Viana, P.~T.~P., Liddle, A.~R., \& Mann, R.~G.\ 2001, \apj~, 547, 594 

\bibitem{} Runyan, M.\ C.\ 2002, PhD Thesis, Caltech.

\bibitem{} Runyan, M.\ C.\ et al.\ 2003, \apj~ in press (astro-ph/0303515).

\bibitem{2003astro.ph..2209S} Spergel, D.~N.~et al.\ 
2003, \apj~ submitted (astro-ph/0302209)

\bibitem{} Sunyaev, R.A. \& Zel'dovich, Ya. B., 1972, Comm. Astrophys. Sp. Phys., 4, 173.

\end{thereferences}
\end{document}